# Molecular dynamics simulation of flow around a circular nano-cylinder

Yanqi Zhu[1], Hanhui Jin[1,2,a)], Yu Guo[1,2], Xiaoke Ku[1,2], Jianren Fan[2]

1. School of Aeronautics and Astronautics, Zhejiang University, Hangzhou 310027, PR China
2. State Key Laboratory of Clean Energy Utilization, Zhejiang University, Hangzhou, 310027, China

a) **Author to whom correspondence should be addressed:** enejhh@emb.zju.edu.cn

**ABSTRACT:** In this study, the wake flow around a circular nano-cylinder is numerically investigated with molecular dynamics simulation to reveal the micro/nano size effect on the wake flow. The cavitation occurring when Reynolds number (Re) > 101 can effectively influence the wake flow. The Strouhal number (St) of the wake flow increases with the Re at low Re, but steadily decreases with the Re after the cavitation appears. The dominant frequency of the lift force fluctuation can be higher than that of the velocity fluctuation, and be drowned in the chaotic fluctuating background of the Brownian forces when Re $\geq$ 127. Also because of the strong influence of the Brownian forces, no dominant frequency of the drag force fluctuation can be observed. The Jz number, which is defined as the ratio between the mean free path λ of the fluid molecules and the equilibrium distance of potential energy $\sigma$, is newly introduced in order to consider the internal size effect of fluid. The St of the wake flow increases with the Jz until it falls to zero sharply when Jz $\approx$ 1.7. It denotes the discontinuity of the fluid can eventually eliminate the vortex generation and shedding. Meanwhile, the St decreases with the Kn because of the intensification of the cavitation.

*Keywords:* molecular dynamics; flow around circular cylinder; vortex shedding; cavitation effect

## I. INTRODUCTION

The flow around a cylinder is a well-known phenomenon in nature, and also is a widely encountered process in engineering. This type of flow field exhibits manifold appearances when the Reynolds number (Re) varies, including vortex motion[1,2], cavitation[3,4], shockwave[5], aeolian tones[6,7], Magnus effect[8], and drag force[9]. Therefore, there have been several experimental and numerical investigations on the process of flow around cylinders. Many experimental studies[10-16] have been conducted to investigate the unique phenomena occurring in the process, including mixed convection heat transfer, buoyancy, wall proximity effects, cylindrical arrays and passive control on the flow response and wake structure. In addition to experimental studies, numerical studies have also been essential in understanding the wake flow, such as the heat transfer of the wake mixing process[17-19], motion of particles in wake[20], effect of cylinder array on flow[15, 21, 22], and the influence of the bluff body shape[23]. The coupling interaction between fluid and the bluff body structure was a key subject in these investigations. For example, the Kármán vortex[24, 25], which usually induces vibration of the structure, was often adapted to characterize the interaction. The flow tended to be stable when Re < 49, and a vortex pair was generated and stayed stationary behind the cylinder. The vortex was shed away from the cylinder periodically when 49 $\leq$ Re $\leq$190. The shedding process can be used to describe the shedding frequency or Strouhal number (St)[26].

Compared to the macro-scale flow, micro/nanoscale flows produce some other non-negligible unique phenomena due to the small size in spatial structure. Such micro/nano size effects can significantly modify the appearance and evolvement of the flow structure, especially when the influence of the discontinuity of the fluid is understood. Rinaldi et al[27] found that the damping, stability, and frequency shift in micro-resonators with internal flow were similar to those in larger resonators. However, Wu et al[28] measured the friction factors of gas flow in the fine channels in microminiature refrigerators, and observed a divergence from classical hydrodynamics theory. Thus, it is necessary to further investigate the micro-size effect on the flow field.

There are also some studies on the flow around a circular cylinder at the micro-scale. For example, Gong et al[29] used the non-equilibrium molecular dynamics (MD) method to simulate high-velocity methane flow in a mica nano slit-pore with various shapes of blockage. A reduced density zone with lower pressure than that of flow in a normal slit-pore was found following the obstacle in their study. Meanwhile, the vortex, reverse flow, and restricted oscillatory were consistent with the flow in macro-volume. Rapaport et al[30, 31] conducted an MD investigation on the two-dimensional flow around a nano circle obstacle. In their study, the flow field was observed to have properties such as the appearance of eddies, periodic eddy separation, and an oscillatory wake, common to real fluids. Sun et al[32] analyzed impeded flow around a single cylinder and two paratactic cylinders at the nanoscale in view of discrete particles; their findings revealed that most flow patterns at the normal scale remained. Although no evident difference was observed in the wake flow of nano-cylinders or nano-obstacles in these studies, recent studies showed that the limited size of the cylinder (or obstacle) could effectively influence the evolvement of the flow field. Asano et al[3, 33, 34] discovered cavitation behind the nano-cylinder in their MD simulation, and they attributed its occurrence to the finite size effect. The St of the Kármán vortex was then calculated with various Re. Unlike in normal wake flows, the nonmonotonic St-Re variation was found in different cases. It is the occurrence of the cavitation that resulted in the nonuniform spatial distribution of the molecule concentration, density, and viscosity behind the nano-cylinder. The disparate findings in previous studies highlight the need for further research on the wake flow around a nano-cylinder and the size effect of the nano-cylinder.

In this paper, the MD method is used to determine the size effect of the nano-cylinder on the wake flow. First, the distinctive flow patterns around a nano-cylinder at various Re are determined. Second, the occurrence of cavitation and its influence on the following wake flow are studied in detail. Third, the size effect is analyzed based on the characteristic cylinder-fluid coupling scale and fluid internal scale.

## II. NUMERICAL SETUP

In this study, the wake flow of argon (Ar) flowing past a silver circular nano-cylinder is adopted. To investigate the microscopic interaction between the cylinder and fluid, the MD method is used so that the discontinuity of the fluid is fully considered. The interaction between the molecules is then represented with the Lennard-Jones potential[35],

$$\phi(r_{ij}) = \begin{cases} \phi_0(r_{ij}) - \phi_0(r_c) & (r_{ij} > r_c) \\ 0 & (r_{ij} \leq r_c) \end{cases} \quad (2.1)$$

$$\phi_0(r_{ij}) = 4\varepsilon\left[\left(\frac{\sigma}{r_{ij}}\right)^{12} - \left(\frac{\sigma}{r_{ij}}\right)^6\right] \quad (2.2)$$

where $r_{ij}$ is the distance between molecules, $\varepsilon$ and $\sigma$ respectively represent energy and the equilibrium distance when potential energy is equal to zero, and $r_c$ is the truncation radius of the potential function. To calculate the balance between the accuracy and efficiency of the computation, $r_c = 3\sigma$ is adopted in present paper. The values of $\varepsilon$ and $\sigma$ depend on the material of the atoms. For the Ar atoms, $\varepsilon_{Ar} = 1.67 \times 10^{-21} J$, $\sigma_{Ar} = 3.4 \times 10^{-10} m$, and atomic mass $m_{Ar} = 6.634 \times 10^{-26} kg$. For the silver (Ag) atoms, $\varepsilon_{Ag} = 5.516 \times 10^{-20} J$, $\sigma_{Ag} = 2.54 \times 10^{-10} m$, and atomic mass $m_{Ag} =$

$1.793 \times 10^{-25} kg$[36]. The interaction between two different kinds of atoms is usually represented by a mixed equation,

$$\sigma_{Ar-Ag} = \frac{1}{2}(\sigma_{Ar} + \sigma_{Ag}), \varepsilon_{Ar-Ag} = \sqrt{\varepsilon_{Ar} * \varepsilon_{Ag}} \tag{2.3}$$

For convenience, the physical variables in the MD simulation are often expressed in the dimensionless MD units, such as energy $\varepsilon$, length $\sigma$, time $\tau = \sigma\sqrt{m/\varepsilon}$ and density $m/\sigma^{dim}$. These parameters are all based on the data of the Ar atom.

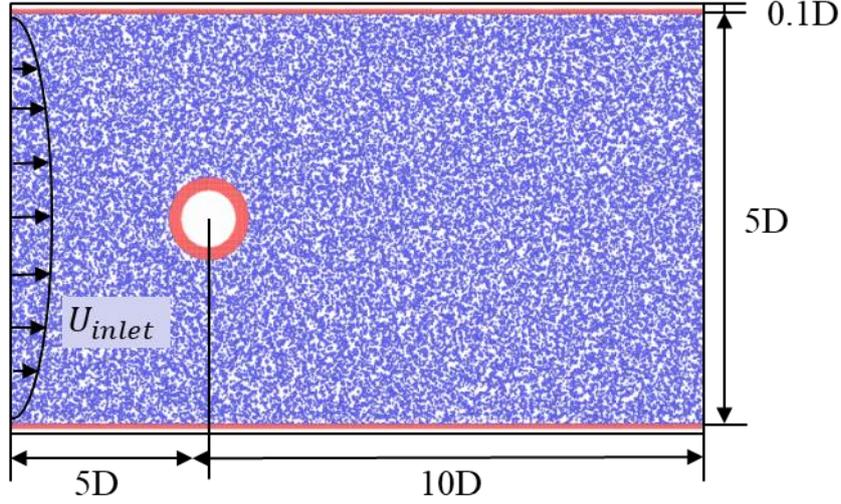

FIG. 1. Schematic diagram of the MD computational zone

Due to limited computation resources, a two-dimensional MD is conducted in a rectangular computational zone with a size of $L_x \times L_y = 600\ nm \times 200\ nm$, namely in MD units of $L_{x,MD} \times L_{y,MD} = 1765 \times 588$ (as shown in Fig. 1). The cylinder with a diameter of $D = 40\ nm$ is placed on the symmetric line with its center 5D distant from the inlet. The lateral and cylinder walls comprise Ag atoms organized in a face-centered cubic arrangement. To reduce the impact of the wall effect, the fully developed parabolic inflow velocity distribution is set at the inlet,

$$U_{inlet,x} = U_0 * \left(1 - \left|2 * \frac{y}{L_y}\right|\right)^2, U_{inlet,y} = 0 \tag{2.4}$$

where $y$ is the transverse distance from the symmetric line, $L_y$ is the width of the computational domain, and $U_0$ is the characteristic maximum velocity. The corresponding Re defined with $U_0$ and D is then obtained as

$$Re = \frac{\rho U_0 D}{\mu} \tag{2.5}$$

where $\rho$ and $\mu$ are the density and viscosity of the fluid at the inlet, respectively. The initial inflow molecules are created in a small periodic flow zone, in which the temperature is set at 130 K and the density is 1300 kg/m³ (MD unit is 0.77). The corresponding viscosity is about $\mu_{Ar} = 1.64 \times 10^{-4} Pa \cdot s$ according to the NIST dataset. About $1.2 \times 10^6$ Ar molecules and $1 \times 10^5$ Ag atoms are involved in the MD simulation.

A GPU accelerated code is developed to perform large-scale MD computation. The implicit velocity-Verlet algorithm is applied to ensure precision. The time step is $\Delta t = 1 \times 10^{-15} s$ (MD unit is 0.00047), and more than $1 \times 10^8$ time steps (about 66 times the cycle period with the mean inflow velocity) are conducted to obtain fully developed results. The time-varying period of wake velocity is selected as a secondary criterion to determine whether the flow is stable.

To determine the spatial distribution and temporal evolvement of the flow field, a pre-set mesh is built over the entire computational zone. The statistically averaged variables of the molecules in each cell are then obtained. Thus, the local statistically averaged velocity $U$ and density $\rho$ are defined as follows:

$$U = \frac{\sum_i^N m_i v_i}{\sum_i^N m_i}, \rho = m_{Ar} * \left(\frac{N}{l_{mesh}^2}\right)^{\frac{3}{2}}, i = 1 \sim N \quad (2.6)$$

where $m_i$ is the mass of $i$ th molecule in a certain cell, $v_i$ is the molecular velocity, $l_{mesh}$ is the mesh size, $N$ is the number of molecules in the mesh cell, and 3/2 is the conversion coefficient from two-dimensional to three-dimensional in the present study.

## III. RESULTS AND DISCUSSION

### A. Cavitation and its influence on the wake flow

Fig. 2 depicts a momentary snapshot of the overall spatial distribution of the vorticity resulting from the statistically averaged velocity field. The shedding of the vortex pair and creation of the Kármán vortex street can be seen. As the Re increases, the shedding frequency of the vortices seems to change and no obvious difference is observed from the normal wake flow around round cylinders.

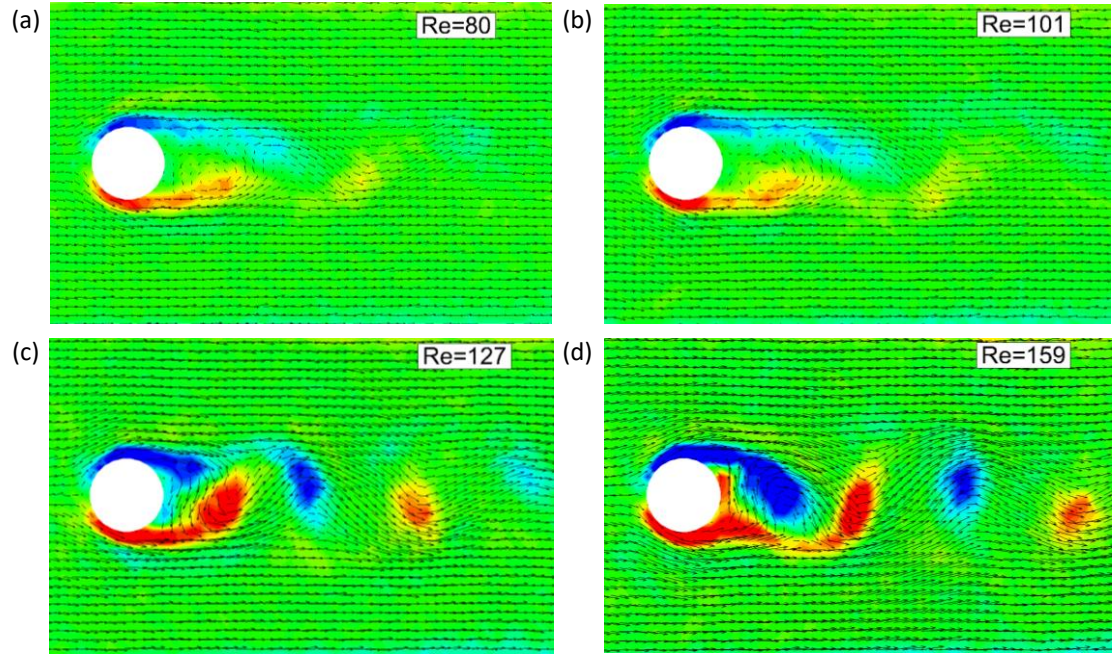

FIG. 2. Spatial distribution of the wake flow field and vorticity contour around the nano-cylinder (a) Re = 80, (b) Re = 101, (c) Re = 127, and (d) Re = 159

Although changes in the vorticity field cannot be observed directly, the distribution of the molecules in the field apparently changes as the Re increases (as shown in Fig. 3). Low concentration of the fluid molecules can be observed behind the cylinder, especially when Re > 127. This refers to the occurrence of cavitation, which was also described in Ref. 33. Meanwhile, the zone of low concentration expands with the Re. Thus, the increasing Re can significantly influence cavitation. When Re < 101, no visible cavitation occurs even though the concentration of the fluid molecules behind the nano-cylinder is much lower than that in other areas. When Re > 127, cavitation is observed behind the nano-cylinder because a gas bubble with very low concentration is present.

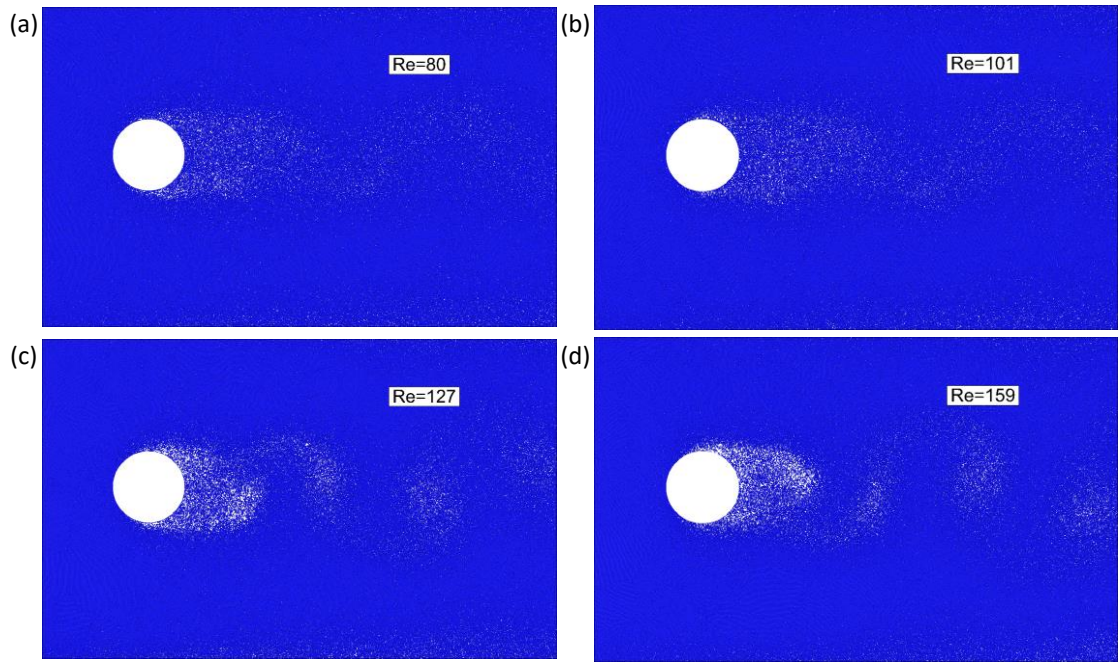

FIG. 3. Spatial distribution of the fluid molecules at different Re (a) Re = 80, (b) Re = 101, (c) Re = 127, and (d) Re = 159

Fig. 4 depicts the temporal velocity fluctuation at different Re behind the cylinder at X/D = 3, where x denotes the streamwise distance from the center of the cylinder. The periodic fluctuation similar to that of normal wake flows is easily observed, except for the burrs near the maximum velocities (as shown in Fig. 4 (a)). The regular fluctuation of the wave amplitude is observed as the Re increases. Further analysis of the frequency spectrum is shown in Fig. 4 (b). The dominant frequency is displayed clearly, and the secondary doubled frequency can also be seen at high Re. Similar to the results obtained in Refs. 29-31, no obvious difference was noted from those in normal cylinder wake flows, except for the different periodic velocity fluctuation waveforms and dominant frequencies at different Re.

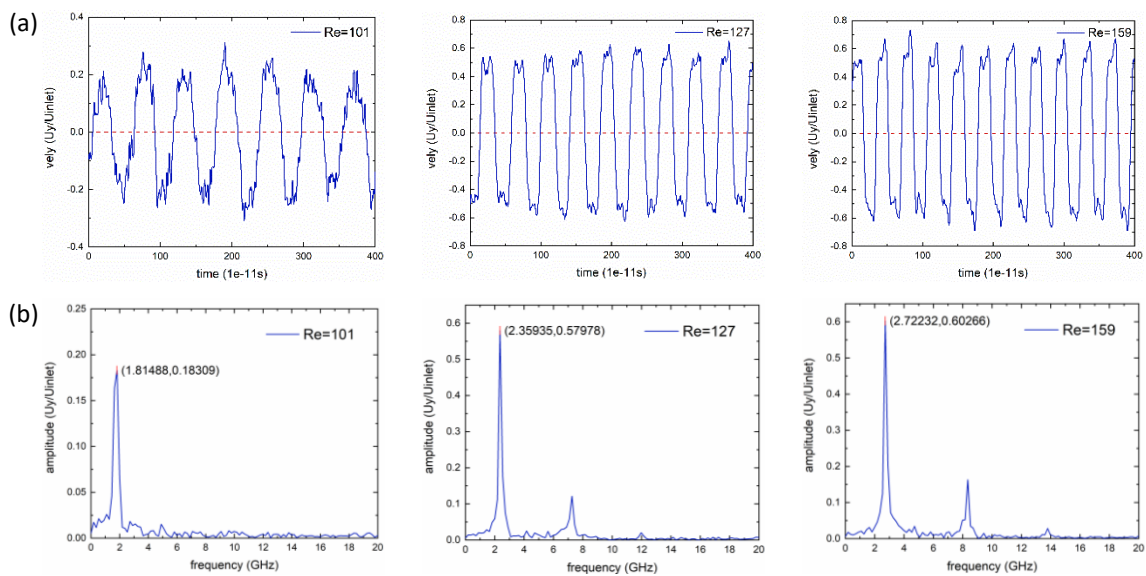

FIG. 4. Periodic velocity fluctuation at wake flow x/D=3 (a) Temporal velocity fluctuation at different Re (b) Frequency spectrum distribution of the velocity

To disclose the variation of the dominant fluctuation frequency with the inflow Re, the St is then utilized

$$St = \frac{Df}{U_0} \qquad (3.1)$$

where $f$ is the dominant frequency of the velocity fluctuation. Fig. 5 shows the St at x/D = 3.0 changing with the inflow Re at the condition of density $\rho = 1300 kg/m^3 (0.77\ in\ MD\ unit)$. Compared with the experimental results obtained by Roshko et al[26] in normal scale wake flows, the nonmonotonic variation of St can be seen in the current nano-cylinder wake flow instead of the monotonic rising. Comparable results were also observed in Asano's MD simulaiton[34]. Although different variation trends were exhibited in different cases in their study, the reversal of the trends above a certain Re was disclosed. To some extent, the agreement of the nonmonotonic variation between the two types of research also proves the validation of MD results in this paper. Such nonmonotonic St-Re variation denotes that the cavitation that occurs in nano-cylinder wake flow substantially changes the variation of St with the inflow Re.

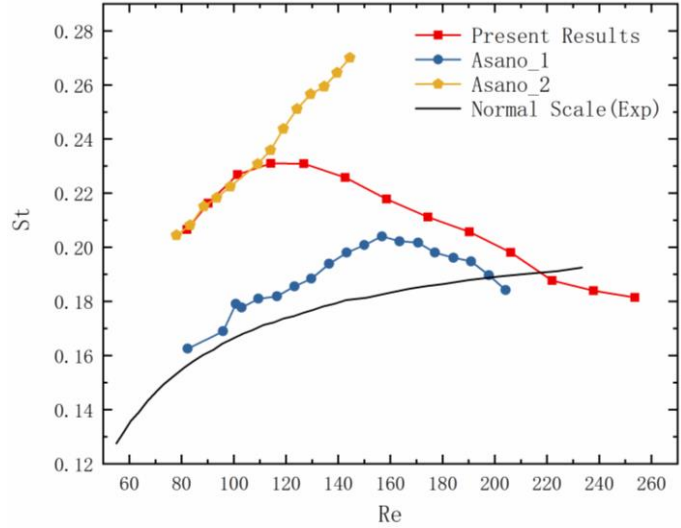

FIG. 5. St-Re variation

(—■— present results $\rho_{MD} = 0.77, L_{y,MD} = 588$, —●— Asano_1[34] $\rho_{MD} = 0.60, L_{y,MD} = 5000$, —●— Asano_2[34] $\rho_{MD} = 0.70, L_{y,MD} = 500$, —— Normal Scale experiment data by Roshko[26])

As shown in Figs. 3 (a) and (b), the decrease in the concentration of molecules is not so intensive at Re < 101, and no evident gas bubble can be observed. This indicates that the cavitation has not taken on the shape yet. Vortex generation and shedding occurs as in regular fluid wake flow so that the St increases with the Re in this phase. As the inflow Re rises, the cavitation occurs more and more intensively. Meanwhile, the gas bubble generated in the cavitation becomes larger and larger (as shown in Figs. 3 and 6). Fig. 6 shows the spatial distribution of the dimensionless density $\rho/\rho_{inlet}$ near the cylinder, in which the shape and coverage of the time-averaged gas bubble zone, transitional zone of gas–liquid two-phase coexistence, and liquid zone are exhibited. It can be noted that the rear tail of the gas bubble zone gets elongated as the inflow of Re increases; meanwhile, the gas bubble zone gradually extends upstream around the nano-cylinder. The appearance of a gas bubble substantially modifies the density of the fluid surrounding the cylinder and related viscosity, which can seriously impact the vortex creation and transit in the wake flow.

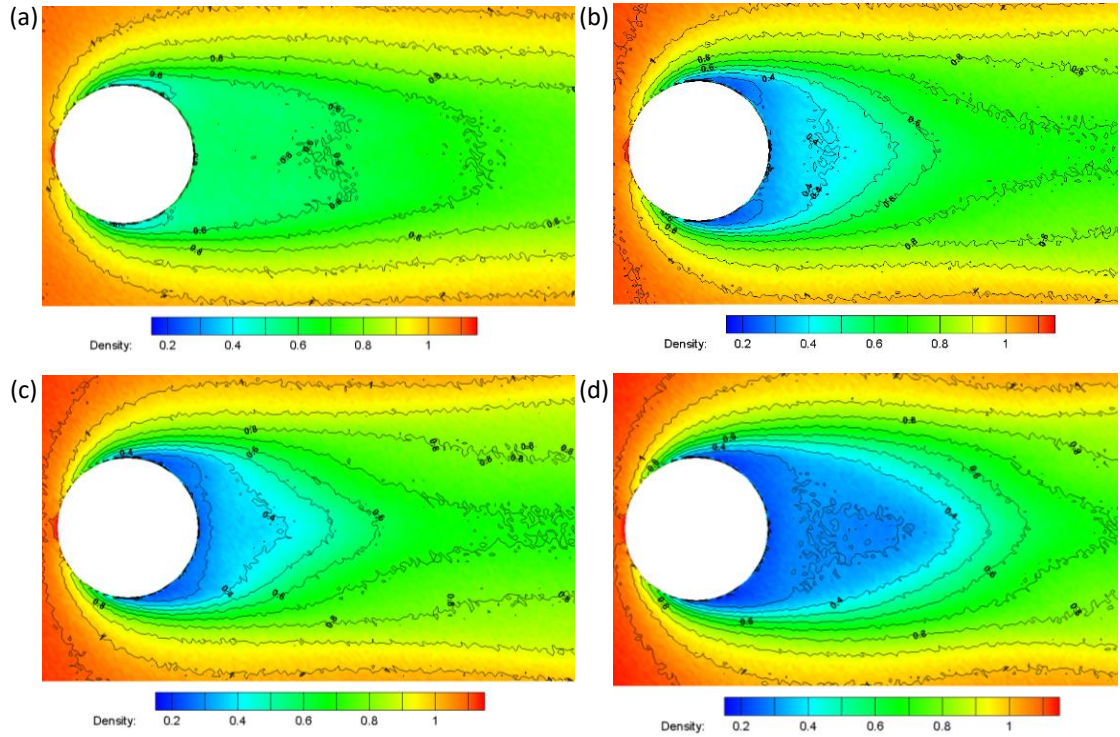

FIG. 6. Spatial distribution of the time-averaged density at different Reynolds numbers (a) Re = 101, (b) Re = 127, (c) Re = 159, and (d) Re = 222

Fig. 7 demonstrates the time-averaged dimensionless density $\rho/\rho_{inlet}$ around the cylinder at a radius of $r = 0.7D$. It exhibits the density behind the nano-cylinder to be primarily $\rho/\rho_{inlet} > 0.5$ while Re < 101 (as shown in Fig. 7 (b)). Corresponding to the St variation in Fig. 5, the vortex shedding essentially coincides with typical wake flow and the St increases with the inflow Re at this phase. When the inflow 101 < Re < 127, the density behind the cylinder decreases quickly. This implies that the emergence of the cavitation and its influence on vortex generation are non-negligible. As a result, the St in Fig. 5 stops increasing with the inflow Re and steadily turns to decrease the inflow Re instead. The density behind the nano-cylinder becomes $\rho/\rho_{inlet} \approx 0.35$ while 127 < Re < 190. At this phase, the density decreases slowly with the Re. This denotes the coexistence of gas-phase and liquid keeps at a relatively stable fraction so that the mode of the vortex street does not alter obviously. It is the transition phase between the non-cavitation and complete cavitation. While Re > 190, the cavitation is intensified seriously to the complete cavitation. The decrease of $\rho/\rho_{inlet}$ is accelerated, and consequently, the gas bubble dominates the zone next to the cylinder. As the inflow Re is improved and the cavitation is correspondingly intensified, the gas–liquid mixing is reinforced so that the length of the mixing zone decreases.

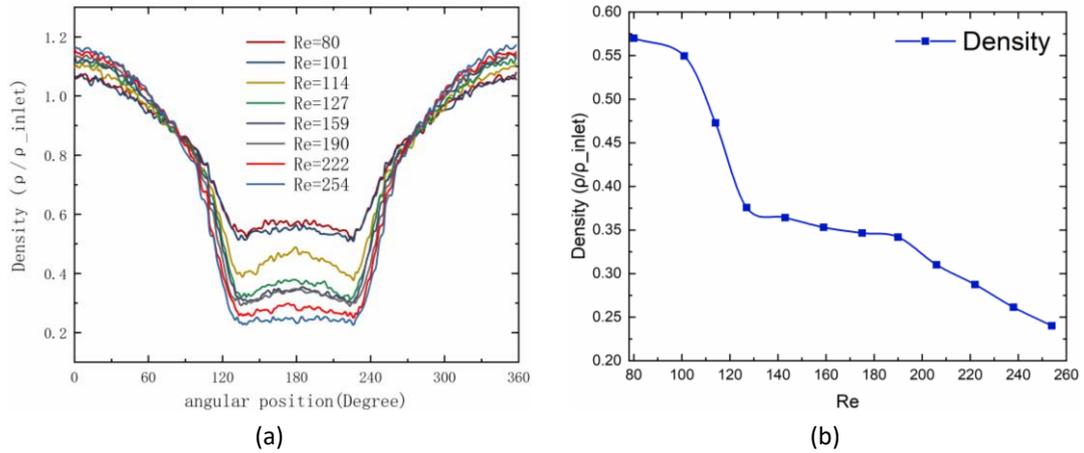

(a)                      (b)

FIG. 7. Density distribution around the nano-cylinder (r=0.7D) at different Re (a) Density distribution around the nano-cylinder, and (b) Minimum density behind the nano-cylinder

If we take the widely used $\rho/\rho_{inlet} = 0.8$ as the critical boundary of the gas–liquid mixing zone, the length of the zone vividly illustrates the multiple cavitation modes at different Re (as shown in Fig. 8). At Re = 127, the mixing length reaches the highest level because of the emergence of significant cavitation and then decreases gradually until Re = 190, where complete cavitation occurs. While the gas bubble is finally created in complete cavitation (Re > 190), the length of the mixing zone remains stable because of the violent mixing and intensive turbulence. The correlation between the figures further reveals that cavitation can substantially influence vortex generation and shedding in the wake flow.

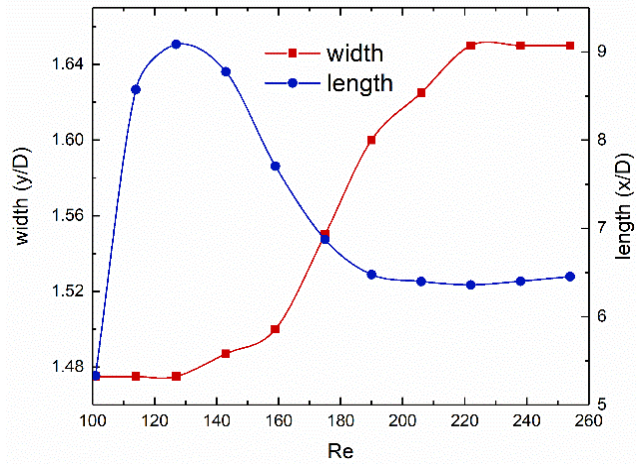

FIG. 8. Width and length of gas-liquid mixing zone at Different Re numbers

Fig. 9 (a) indicates that vortex generation occurs in a region of low molecule concentration, such as in the gas bubble region while the cavitation occurs. It denotes that the vortex is generated in the gas bubble instead of the liquid at the inlet in the presence of cavitation. Thus, the fluid in the cylinder wake flow field has essentially changed from liquid to gas. Correspondingly, the variation in density and viscosity necessarily alters vortex generation and shedding. Meanwhile, the relative velocity between the gas flow and nano-cylinder comes to be the effective inflow velocity of the cylinder wake flow. Because of the slip between the gas and liquid phase, the effective Re becomes significantly less than the inflow Re so that the vortex shedding frequency in Fig. 5 appears to drop with the inflow Re. The fall of effective Re is also illustrated by the peak velocity in the boundary layer around the cylinder. As shown in Fig. 9 (b), the peak velocity exhibits the largest value at Re = 127 and subsequently declines with the inflow velocity. Considering that the peak velocity arises in the gas bubble where the vortices are created, it can be concluded that the effective Re at Re > 127 is lowered by the occurrence of cavitation despite the increase of the inflow Re.

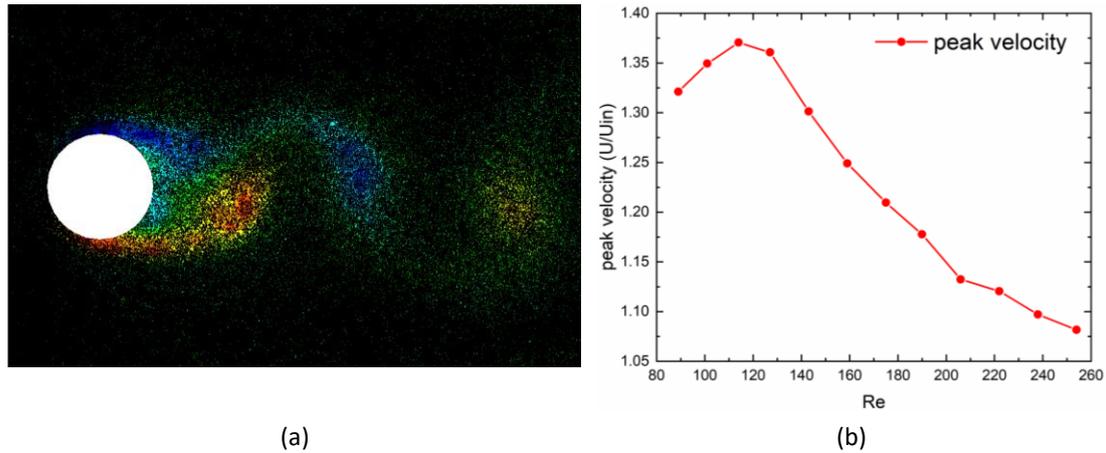

(a)            (b)

FIG. 9. (a) Spatial distribution of the fluid molecules and vorticity contour near cylinder, and (b) peak velocity of boundary layer with different Reynolds numbers

Fig. 10 demonstrates the spatially fluctuated density and velocity in the streamwise direction at a certain moment. It can be observed that the fluctuation frequencies of the streamwise velocity $U_x$ and the density are the same, whereas the fluctuation frequency of the transverse velocity $U_y$ shows double them. Such an interesting phenomenon results from the gas–liquid mixing process, and the fluctuation amplitude of $U_y$ depends on the mixing liquid velocity.

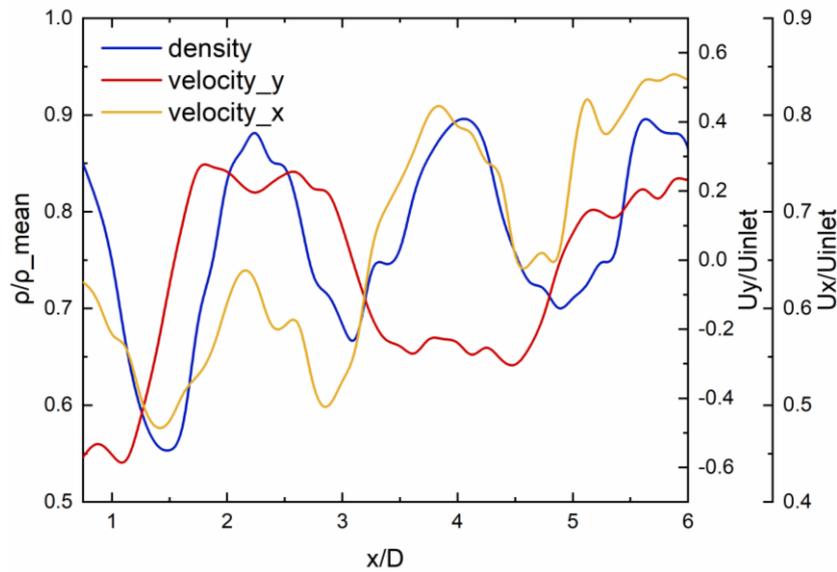

FIG. 10. Transient variation of the velocity and the density distribution in streamwise direction (at $t = 2 \times 10^{-8} s$)

Figs. 11 and 12 depict the frequency spectrum of the lift force and drag force acting on the nano-cylinder, respectively. Although the periodic fluctuation with the dominant frequency of the velocity near the nano-cylinder is visible, it is difficult to determine the dominant frequency of the lift force and drag force when the inflow Re is high enough. As shown in Fig. 11, an obvious dominant frequency of lift force can only be observed when Re < 101. Nevertheless, the dominant frequency at Re = 101 floats away from the dominant frequency of the velocity fluctuation; this is because of the simultaneous fluctuation of velocity and density resulting from the occurrence of the cavitation. Meanwhile, the chaotic background fluctuation is intensified when contrasted with the velocity fluctuation. It can result from the small size of the nano-cylinder and the highly nonuniform distribution of fluid density. The Brownian force can no longer be disregarded in this circumstance. As the inflow Re increases, the cavitation is correspondingly intensified, and the lift force is gradually weakened so that it is drowned in the chaotic background of the Brownian forces. For the drag force, no apparent dominant fluctuation frequency can be obtained because of the large difference in density before and behind the cylinder (as shown in Fig. 12).

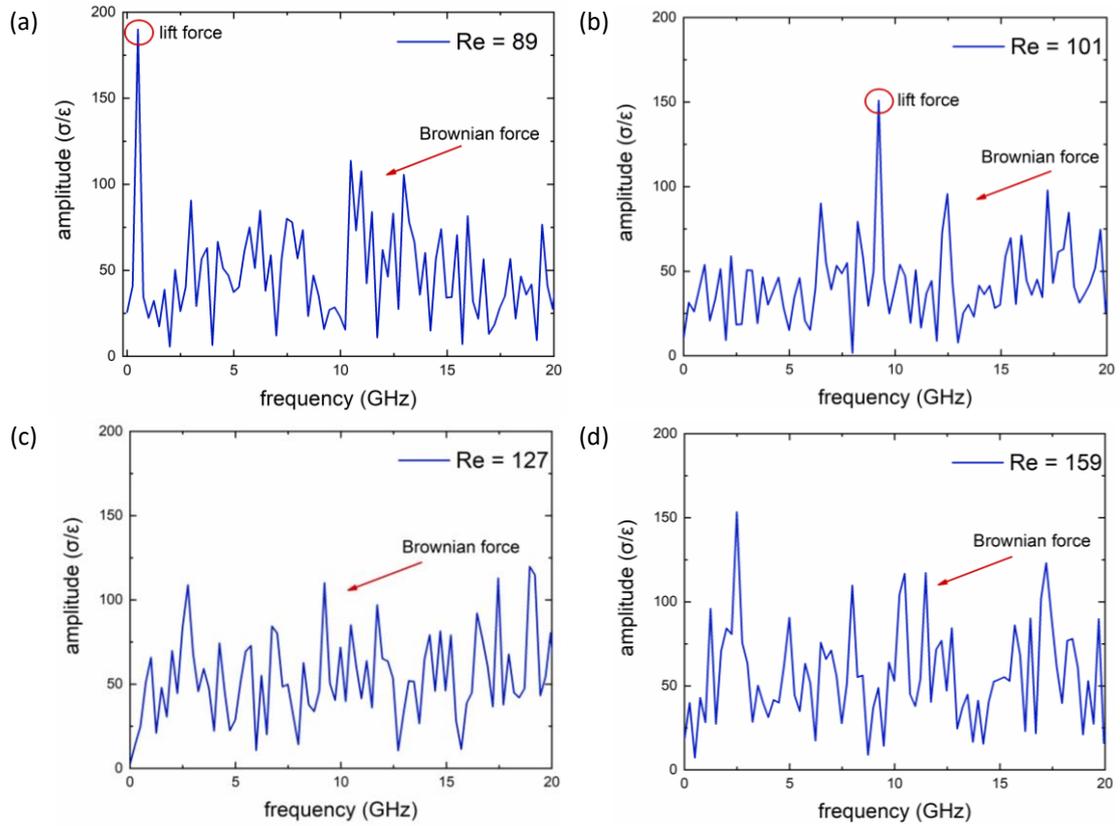

FIG. 11. Frequency spectrum distribution of the lift force at different Re numbers (a) Re = 89, (b) Re = 101, (c) Re = 127, and (d) Re = 159

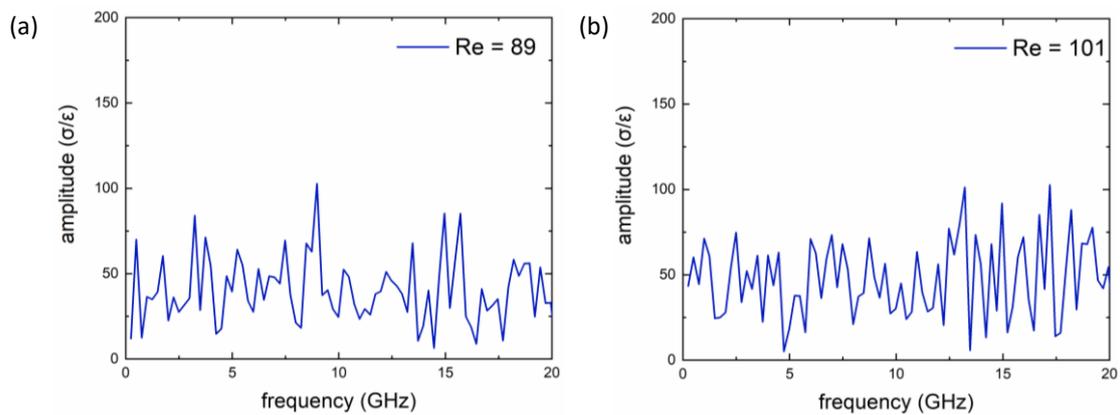

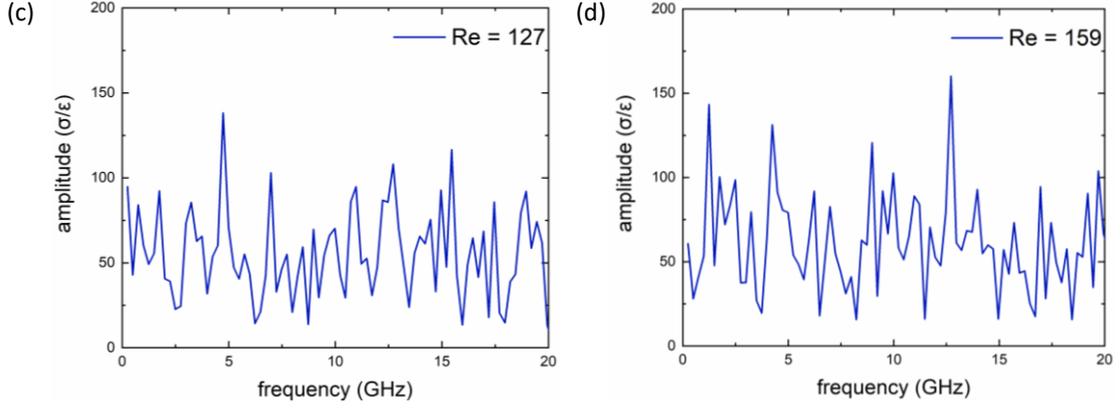

FIG. 12. Frequency spectrum distribution of the drag force at different Re numbers (excluding the mean force) (a) Re = 89, (b) Re = 101, (c) Re = 127, and (d) Re = 159

### B. Scale effect of the interactive size ratios

Traditionally, a critical point[37] of the Knudsen number Kn = 0.01 is commonly adopted to distinguish the flow qualities between the non-slip and slip flow, or between continuous and non-continuous flow. Because of the limited size of the nano-cylinder wake flow, the trans-scale coupling between the microscopic sizes such as the inter-molecule distance and the characteristic scales of the wake flow such as the diameter of the nano-cylinder can essentially influence the flow pattern of the wake flow field. Considering the trans-scale effect, two critical size scale ratios are chosen in the present paper. The first one is Kn, which is the ratio between the mean free path of the fluid molecules λ and the characteristic length of the flow field, which is designated as the nano-cylinder diameter D in the present paper, namely

$$Kn = \frac{\lambda}{D}, \lambda = \frac{1}{\sqrt{2}\pi d^2 n} \tag{3.2}$$

where $D$ is the diameter of the cylinder on behalf of the characteristic length of the flow field, $d$ is the effective molecular diameter, and $n$ is the molecular number density. The other one is the $Jz$ number, which is a newly introduced parameter in the present paper. The Jz number is defined as the ratio between the mean free path λ and the equilibrium distance of potential energy σ

$$Jz = \frac{\lambda}{\sigma} \tag{3.3}$$

$Jz$ is a parameter to characterize the interaction intensity between molecules. If the $Jz$ is large enough, the mean interactive forces between molecules would be very small. Thus, the molecules of the fluid can move freely without any apparent limitation from other molecules, and the fluid then turns to be non-continuous or compressible. Kn denotes the continuity of the interaction between the fluid and nano-cylinder, and $Jz$ denotes the continuity of the interaction between the internal molecules of the fluid that can influence the flowability. The two parameters potentially determine the flow pattern of the wake flow in the way of static physical property. For a given cylinder wake flow, the two parameters usually grow together.

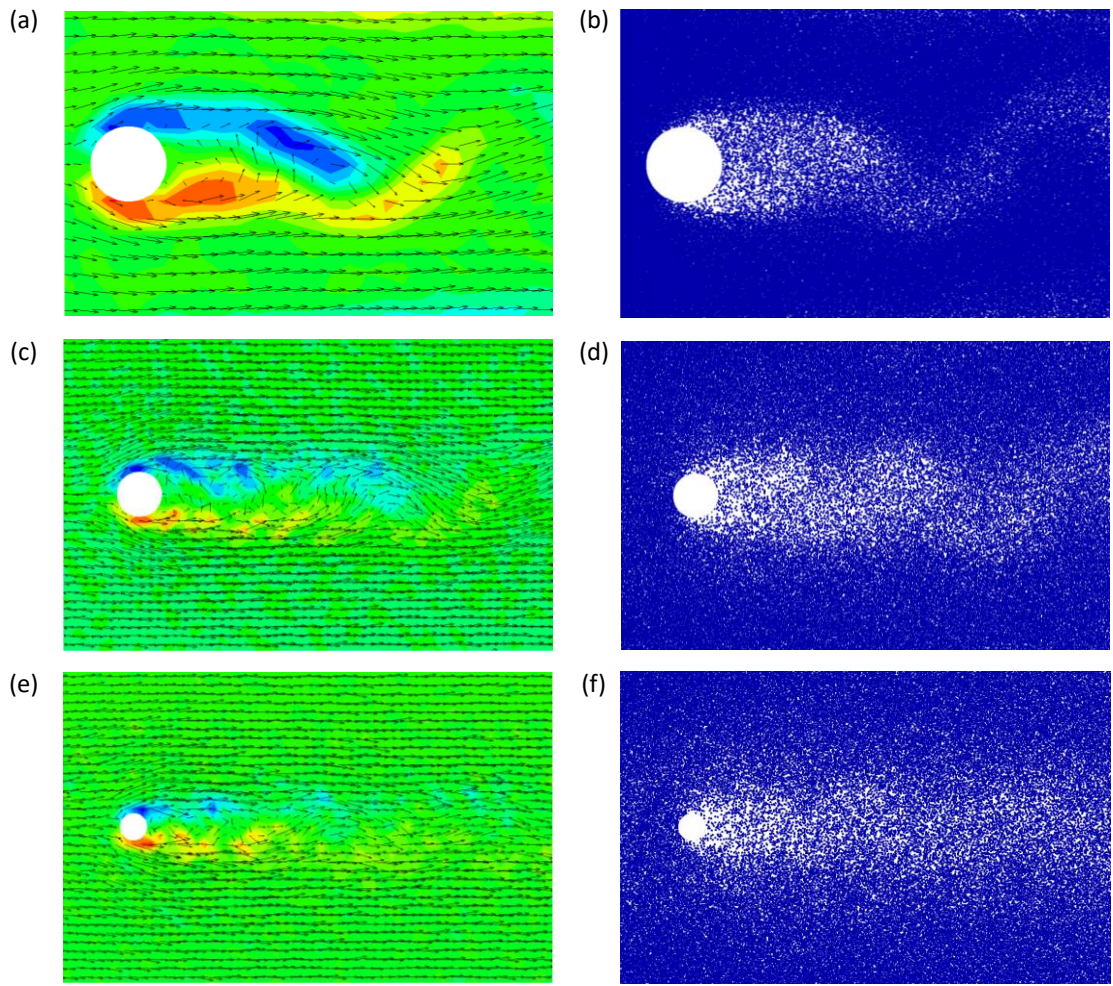

FIG. 13. Spatial distribution of the wake flow field and the fluid molecules at different Kn numbers (Re =127 and Jz =0.195)

(a) Vorticity contour and velocity vectors at Kn = 0.00331, (b) Spatial distribution of molecules at Kn = 0.00331, (c) Vorticity contour and velocity vectors at Kn=0.00662, (d) Spatial distribution of molecules at Kn=0.00662, (e) Vorticity contour and velocity vectors at Kn=0.011, and (f) Spatial distribution of molecules at Kn=0.011

Fig. 13 displays the flow patterns at various Kn with the fixed Jz = 0.195 and Re = 127. As the Kn increases, the configuration of the vortex roadway can no longer be illegible. Nevertheless, the nonuniform distribution of the fluid molecules resulting from the cavitation is exacerbated with the Kn. Despite the difference between different Kn, the basic configuration of the wake flow does not change considerably. A similar phenomenon can also be observed for various Jz with the fixed Kn = 0.0108 and Re = 127 (as shown in Fig. 14).

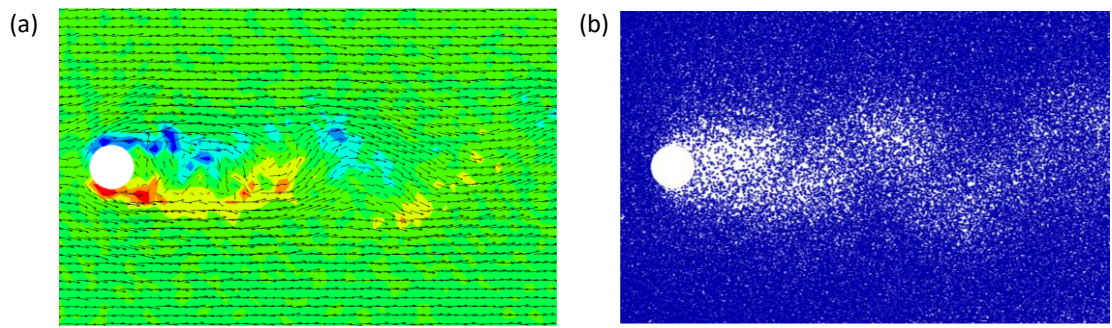

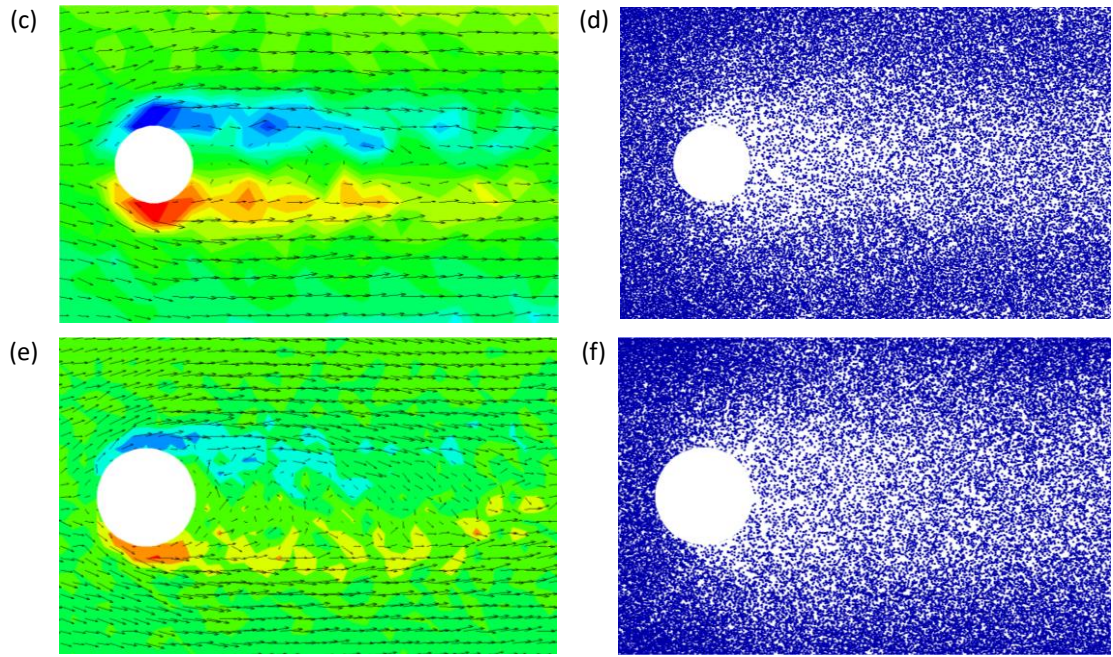

FIG. 14. Spatial distribution of the wake flow field and the fluid molecules at different Jz numbers (Re =127, Kn =0.0108)
(a) Vorticity contour and velocity vectors at Jz=0.253, (b) Spatial distribution of molecules at Jz=0.253, (c) Vorticity contour and velocity vectors at Jz=0.633, (d) Spatial distribution of molecules at Jz=0.633, (e) Vorticity contour and velocity vectors at Jz=1.27, and (f) Spatial distribution of molecules at Jz=1.27

Fig. 15 shows the distribution of the transverse velocity in different cross-sections. The non-smooth variation of the velocity can be observed near the gas–liquid interface in Figs. 15 (a) and (b). It denotes the occurrence of the phase transition in the cavitation and the changes in fluid physical parameters, including viscosity and density, near the interface. However, such variation is not visible in Fig. 15 (c). It means no visible phase transition contact exists near the nano-cylinder while Jz = 2.53 and Kn = 0.0215. Thus, no cavitation occurs in this scenario, and the decrease in density behind the cylinder occurs because of the compressibility of the rarefied gas.

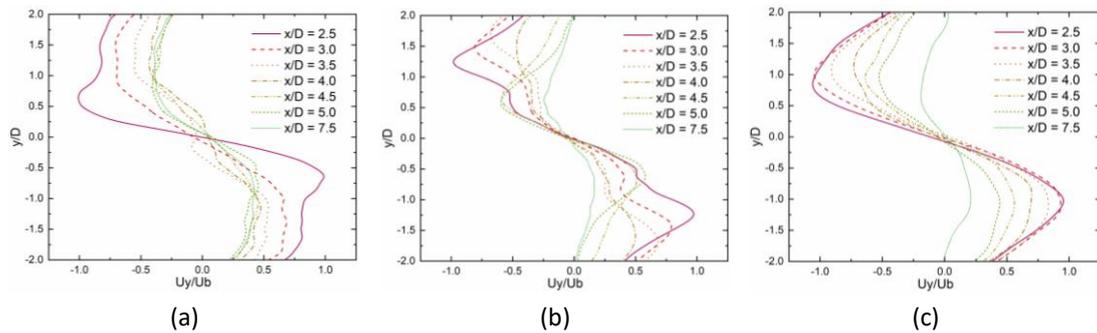

FIG. 15. Distribution of the transverse velocity in different cross-sections (a) Kn = 0.00166, Jz=0.253, (b)Kn=0.00538, Jz=0.633, and (c)Kn=0.0215, Jz=2.53

Fig. 16 illustrates the frequency spectrum of the velocity fluctuation at different Kn and Jz when Re = 127. It can be easily observed that the dominant frequency varies with the Kn and Jz. Although the variation trends of the dominant frequency cannot be obtained directly from the figure, the disappearance of the dominant frequency can be observed when the Kn or Jz increases to a certain value.

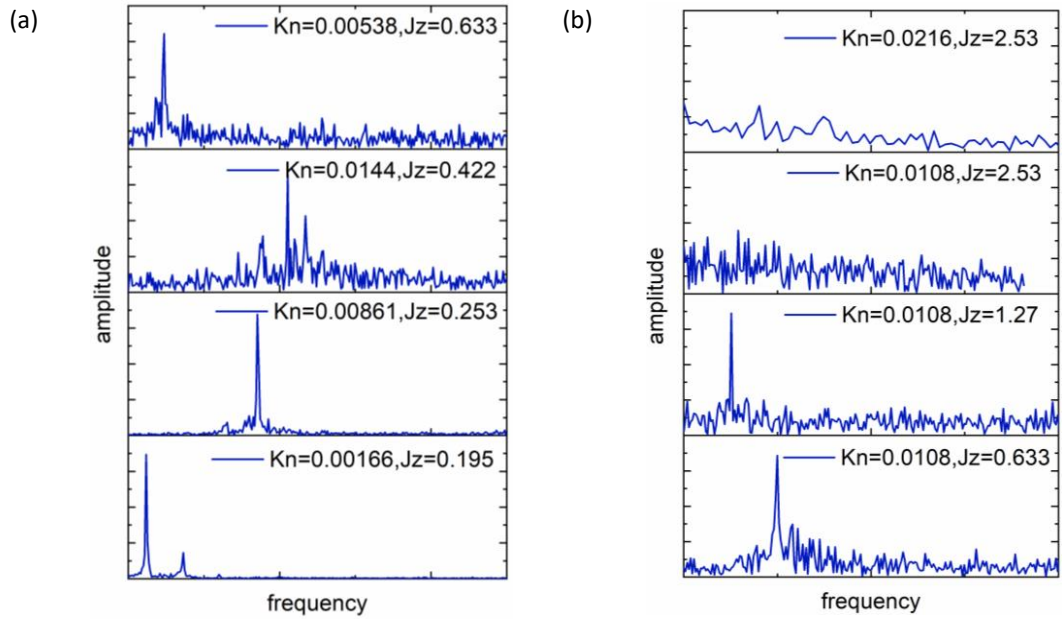

FIG. 16. Frequency spectrum distribution of the velocity fluctuation at different scale ratio (Kn and Jz)

To disclose the detailed influence of the Kn and Jz on the wake flow, the St-Kn and St-Jz variations at Re = 127 are obtained in Fig. 17 together with the minimum dimensionless density $\rho/\rho_{inlet}$ behind the cylinder. The Kn and Jz have a different impact on the St. The St decreases with the Kn but increases with the Jz. Improving the Kn can intensify the cavitation in the wake flow so that the fluctuation frequency of the velocity is decreased. Such variation is similar to the increase of the Re while Re > 127 in Fig. 5. However, improving the Jz at a fixed Kn can decrease the surface tension near the gas–liquid interface, which can intensify the mixing between the gas and liquid phase. The intensification of the gas–liquid mixing inhibits the development of the cavitation, and consequently, the St number is improved. Meanwhile, it is also noted that the St drops dramatically when Kn > 0.15 or Jz > 1.27; this result denotes that the discontinuity of the fluid can eventually abolish the vortex generation and shedding in the wake flow.

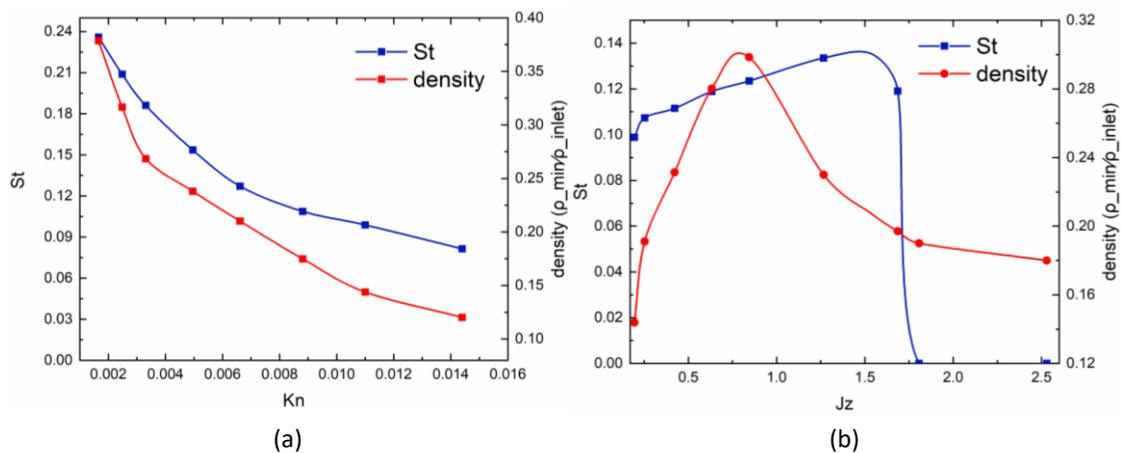

(a)　　　　　　　　　　　　　　　(b)

FIG. 17. Effect of scale ratios on the St and the lowest density behind the nano-cylinder

(a) Effect of Kn at Jz = 0.195, and (b) Effect of Jz at Kn = 0.0108

## IV. CONCLUSIONS

In this study, MD modeling of the wake flow around a circular nano-cylinder is conducted and the micro/nano size effect on the wake flow is numerically explored. The occurrence of cavitation and its effect on the flow pattern, lift force, and drag force are explored at different inflow Re. Meanwhile, the scale effect of the size ratios, such as the Kn and Jz, is also revealed.

1. While Re < 101, no discernible difference can be observed between the nano-cylinder wake flow and the normal wake flow. However, when Re > 127, the occurrence of cavitation can dramatically alter the wake flow, particularly the frequency of vortex generation and shedding from the nano-cylinder. The St grows with the Re at low Re but gradually decreases as the Re reaches a critical value.
2. Because of the concurrence of the density and velocity fluctuation caused by the cavitation, the dominant frequency of the lift force fluctuation can be higher than that of the velocity fluctuation. The Brownian forces are the dominating forces on the cylinder so that the dominant frequency of the lift force is drowned in the chaotic fluctuating background of the Brownian forces when Re ≥ 127. In addition, because of the substantial influence of the Brownian forces, no dominant frequency of the drag force fluctuation can be observed, except for the mean drag force.
3. The Jz number, which is defined as the ratio between the mean free path λ of the fluid molecules and the equilibrium distance of potential energy σ, was recently developed to consider the internal size effect of fluid. Jz can determine the compressibility and continuity of the fluid. Our results reveal that the increase in Jz can significantly improve the gas–liquid mixing intensity in the cavitation and correspondingly inhibit the formation of the cavitation. The St of the wake flow increases with the Jz when Jz < 1.27 but decreases rapidly when Jz > 1.27 because of the discontinuity of the fluid.
4. The increase in Kn can decrease the fluid density behind the cylinder and intensify the cavitation. The St decreases with the Kn when Kn < 0.15 and decreases even faster when the discontinuity of the connection between the fluid and cylinder dominates at Kn > 0.15. The variation of St at high Kn or Jz denotes that the discontinuity of the fluid can finally abolish the vortex generation and shedding in the wake flow.

## ACKNOWLEDGEMENTS

The authors gratefully acknowledge the financial support of the Natural Science Fund of China (Grant No: 91741103).

## AUTHOR DECLARATIONS

### Conflict of Interest

The authors declare no conflicts of interest.

## DATA AVAILABILITY

The data that support the findings of this study are available from the corresponding author upon reasonable request.

---